# Telecom C-Band Photon-Pair Generation using Standard SMF-28 Fiber

Kyungdeuk Park[a], Dongjin Lee[a], Robert W. Boyd[b,c], Heedeuk Shin[a,*]

[a]Department of Physics, Pohang University of Science and Technology (POSTECH), Pohang 37673, Korea
[b]Department of Physics, University of Ottawa, 25 Templeton, Ottawa, Ontario, K1N 6N5, Canada
[c]Institute of Optics, University of Rochester, Rochester, New York 14627, USA
*heedeukshin@postech.ac.kr

**Abstract:** Photon-pair generation must satisfy both the energy conservation and phase-matching conditions with a specific pump wavelength and dispersion of nonlinear optical medium, but finding a photon-pair, which has a desired specific wavelength, generation medium is challenging. Here, we present a method to create photon pairs that functions efficiently even the pump wavelength is much larger than the zero GVD wavelength of medium. In this study, we employ short SMF-28 fibers having ~1310 nm zero GVD wavelength and C-band pump (1552.52 nm) to generate C-band photon pairs. The measured pair generation rate and coincidence-to-accidental ratios are comparable to those from a long dispersion-shifted fiber. Polarization-entangled states are prepared, and an S value of 2.659±0.094 is achieved from Bell inequality measurements. Our results indicate that the use of a short SFWM medium yields adequate photon-pair generation rates regardless of its dispersion properties in almost any material at any pump wavelength.

## 1. Introduction

A robust and efficient non-classical light source is one of the essential components for the emerging field of quantum-enhanced technologies such as quantum computing [1], quantum sensing [2], and quantum cryptography [3]. Quantum light sources using photon pairs generated by spontaneous parametric down-conversion (SPDC) or spontaneous four-wave mixing (SFWM) have made significant contributions to the development of quantum optics experiments for decades. Unlike SPDC, which requires non-centrosymmetric material [4], the SFWM process is a third-order nonlinear process, and thus even amorphous media such as optical fibers can serve as a photon-pair source. Most importantly, as this photon-pair source based on the use of an optical fiber, it can be coupled with minimum loss to another fiber based quantum optical system, such as a quantum cryptography system.

In the standard SFWM process, the wavelengths of the generated photons are strongly related to the pump wavelength ($\lambda_p$) and dispersion properties of the medium [5]. Wavelengths of wavevector mismatch $\Delta k = 0$ (phase matchting wavelength of generated photon-pair), where SFWM occurs strongly, are close to $\lambda_p$ in the anomalous GVD region (referred to as the fundamental modes) and are getting closer to $\lambda_p$ as the $\lambda_p$ is further away from the zero group-velocity-dispersion (GVD) wavelength ($\lambda_{ZGVD}$), the generated photon-pair, therefore, becomes difficult to separate from pump. And the phase matching wavelengths are far from $\lambda_p$ in the normal GVD region (referred to as the side modes), and are changed greatly even if the $\lambda_p$ is changed a little. For this reason, various types of SFWM media have been used, including dispersion-shifted fiber (DSF) [6-9], photonic crystal fiber (PCF) [10-15], birefringent fiber [16-17], chalcogenide fiber [18], tapered micro/nano-fiber [19-20], and on-chip devices with various structures and materials [21-29]. To generate efficiently photons at desired wavelengths, it is important to select a medium with the proper $\lambda_{ZGVD}$, but finding such a medium is sometimes not simple. In the fundamental mode, to separate pair form pump, pair generation bandwidth is important as well as their phase matching wavelength. If the spectral bandwidth of pair generation can be sufficiently broadened, photon-pair can be well separated from pump regardless of $\lambda_{ZGVD}$, and such a source would be practically useful for the quantum light source of quantum information processing and quantum optics experiments.

Here, we present a straightforward and practical method to generate photon pairs regardless of the dispersion properties of the SFWM medium. It is well known that if the length of the SFWM material, $L$, is short, the phase mismatch parameter, $\Delta k L$, becomes small, yielding the broad pair-generation rate spectrum [13]. Even though this spectral broadening due to short medium length have been reported in the case of on-chip straight waveguide and short specific highly non-linear fiber, like PCF, but, to our best knowledge, has not been experimentally investigated yet in the case that the $\lambda_p$ is much larger than $\lambda_{ZGVD}$. We employ short single-mode fibers (SMF) (Corning, SMF-28) as the SFWM medium to test our method. The pump wavelength ($\lambda_p$=1552.52 nm) is over 240 nm apart from $\lambda_{ZGVD}$ of SMF (~1310 nm). The coincidence counts and coincidence-to-accidental ratio (CAR) are measured for various fiber lengths, showing substantial numbers of photon pairs and high CAR values. In addition, we perform Bell's inequality experiments using polarization-entangled photon pairs, showing good polarization entanglement. The results in this study show sufficiently equivalent properties to those of previously reported results which is from the constraint case that the $\lambda_p$ is near $\lambda_{ZGVD}$, and indicate that photon-pair generation without restriction on $\lambda_{ZGVD}$ selection will be of great benefit for quantum communication applications from a practical point of view.

## 2. Theory of pair-generation rate with phase mismatch parameter

Like the SPDC process, SFWM pair generation requires energy conservation and phase-matching conditions. The wavelengths of created pairs are dominantly determined by the dispersion of SFWM medium as well as pump wavelength and slightly variable by pump power due to the self-phase modulation [5]. The photon-pair generation rate (PGR) can be described as [25, 30]

$$\text{PGR} = \mu_p P_p^2 L^2, \tag{1}$$

where $\mu_p$ is defined as the pair-generation coefficient ($\mu_p \equiv \tau_P f_p B \gamma^2 \text{sinc}^2[\Delta k(\Delta \nu) L/2]$), $\tau_P$ is pump-pulse duration, $f_p$ is pump-repetition rate, $B$ is signal/idler filter bandwidth, $\gamma$ represents the nonlinear Kerr coefficient of the medium, $\Delta k(\Delta \nu) = 2k(\omega_p) - k(\omega_p + 2\pi\Delta\nu) - k(\omega_p - 2\pi\Delta\nu) - 2\gamma P_p$ is the wavevector mismatch between the pump and signal/idler photons for degenerate pumps, $\Delta\nu$ is frequency detuning from the pump frequency, $P_p$ is the pump peak power, and $L$ is the SFWM medium length. The frequency of the pump, signal, and idler photons satisfies the law of conservation of energy. Fig. 1. (a) represents the GVD of SMF-28 fiber against the pump wavelength. Fig. 1. (b) and (c) shows the wavevector-mismatch function and normalized $\mu_p$ spectrum of the SMF against the wavelength difference between the generated photons and the pump photons. The greater the GVD values in Fig. 1. (a) are, the faster the wavevector-mismatch function in Fig. 1. (b) increases. The normalized $\mu_p$ spectrum, as seen in Fig. 1. (c) gets narrower as the pump wavelength is farther from the zero GVD wavelength. Without the careful selection of the zero GVD wavelength with respect to the pump wavelength, it makes it challenging to distinguish between the pump light and created two-photon states.

In Eq. (1), the PGR spectrum has a shape of $\text{sinc}^2$-function whose bandwidth is determined by the phase mismatch parameter $\Delta k L$. At near $\lambda_{ZGVD}$, $\Delta k$ increases slowly as the wavelengths of signal and idler photons become far from $\lambda_p$. Therefore, the PGR spectrum has a broad bandwidth even with a long SFWM medium. However, if $\lambda_p$ is far from $\lambda_{ZGVD}$, $\Delta k$ boosts rapidly as the frequency detuning (Δν) increases, yielding narrow bandwidth of the PGR spectrum. If the length of the SMF-28 fibers is over hundreds of meters like the typical SFWM experiments, the PGR spectrum is narrow, so that it is challenging to separate signal and idler photons from the pump light. The short length of SMF-28 fibers, however, reduces the phase mismatch parameter $\Delta kL$, yielding a broad bandwidth of the pair-generation coefficient spectrum. Then we can readily separate the generated photons from pump light using commercially available filters. The maximum PGR decreases by shortening the medium length, but can be compensated by increasing $P_p$ as the PGR is proportional to both $P_p^2$ and $L^2$ which is well known property.

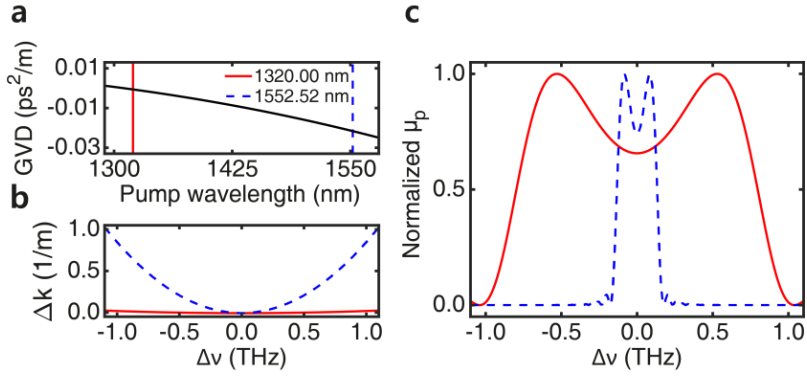

Fig. 1. (a) The GVD of the SMF-28 fiber with the dispersion of 69.7 s/m³ and the zero GVD wavelength of 1310 nm, respectively. (b) and (c) represent the phase-mismatch function ($\Delta k$) and the normalized of the SMF-28 fiber against the wavelength difference between the generated photons and the pump photons. The red and blue lines indicate the pump wavelength of 1320 nm and 1550 nm, respectively. The pump peak power is 3 W and the fiber length is 308 m.

Fig. 2. (a) shows the calculated pair-generation coefficient, $\mu_p$, spectrum using Eq. (1) as a function of the frequency detuning from pump frequency (193.1 THz, equivalent to $\lambda_p$=1552.52 nm). $\Delta k$ is estimated from the dispersion properties of the SMF-28 datasheet [31]. The lengths of fibers are 3.8 m (solid red line), 11.4 m (dashed green line), 31.5 m (dotted blue line), and 308 m (dash-dotted dark-yellow line) at 3-W pump peak power. As expected, the shorter fiber length is, the broader the pair-generation coefficient spectrum is. For a 3.8-m fiber, the spectrum in Fig. 2. (a) is so broad that photon pairs can be generated even at 1000-GHz frequency detuning out of the pump frequency. On the other hand, the bandwidth of a 308-m fiber gets so narrow that the SFWM photons can hardly be detected even at the 200-GHz separated channel. The maximum $\mu_p$ in Fig. 2. (a) is the same as the PGR spectrum is normalized by $P_p^2$ and $L^2$.

Fig. 2. (b) shows the estimated half-width half-maximum (HWHM) bandwidth of the PGR spectrum against the fiber length. The HWHM bandwidth decreases as increasing the fiber length like SPDC [32]. For a 3.8-m fiber, the bandwidth of photon pair generation is about 800 GHz, but the photons from a 308-m fiber can hardly transmit the 200-GHz separated filter set from the pump due to the narrow pair-generation bandwidth (~125 GHz), resulting in negligible detection of signal and idler photons. Furthermore, the 308-m-fiber spectrum in Fig. 2. (a) shows a peak corresponding to the frequency difference between pump and phase-matching ($\Delta k$ =0) frequencies (blue dashed line in Fig. 2. (b)), and this separation behavior is caused by the nonlinear contribution of the pump power to the phase-matching condition ($2\gamma P_p$). The pump peak power is 3-W, showing a peak separation of 77.4 GHz (~0.62 nm).

As mentione above, the reduced PGR can be compensated by increasing pump power, the spectrum, however, should not be changed. Actually, the pump peak power affects to the maximum $\mu_p$ (proportional to the $P_p^2$) and also the bandwidth of spectrum because the $\Delta k$ is changed by self-phase modulation ($2\gamma P_p$). Fig. 2. (c) is the pair generation coefficient ($\mu_p$) for the various $P_p$. The pump peak powers are 1 W (red solid line), 5 W (green dashed line), 10 W (blue dotted line), 50 W (dark yellow dash-dotted line), and 100 W (violet dash-double dotted line) with $L = 3.8\ m$. The $\mu_p$ is not significantly affected by $P_p$ up to 10 W, and is slightly changed after 10 W. Actually, this changes are shift of phase-matching frequency caused by self-phase modulation ($2\gamma P_p$), and the bandwidth does not change much. Therefore, since the contribution of $P_p$ to the bandwidth is small, increasing $P_p$ can be used for compensating maximum $\mu_p$ decrease without changing the bandwidth of spectrum for a short fiber.

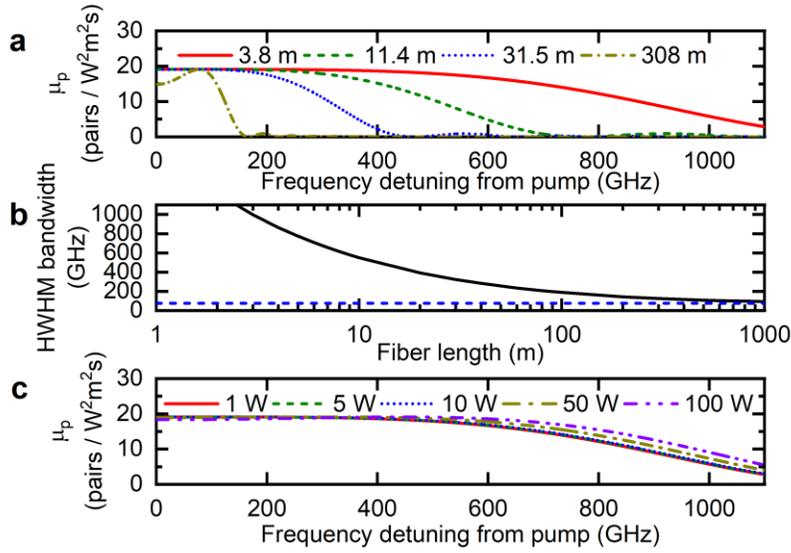

Fig. 2. (a) The pair-generation coefficient $\mu_p$ for various fiber lengths. (b) Estimated half width at half maximum (HWHM) bandwidth of the $\mu_p$ spectrum (Black curve) against the fiber length. The blue dashed line indicates the frequency difference between pump and phase-matching frequencies at the given pump peak power assuming an infinitely long fiber. The horizontal axis is plotted on a logarithmic scale. (c) The $\mu_p$ for various pump peak powers with a short fiber ($L = 3.8\ m$).

## 3. Experimental measurements

In order to verify the generation of telecom C-band (1530 ~ 1565 nm) photon pairs from SMF-28 fibers, we measured the coincidence counts for various fiber lengths [3.8 m, 5.8 m, 8.1 m, 11.4 m, 31.5 m, 55.5 m, 104 m, and 308 m], using the experimental setup in Fig. 3. (a). The length of fibers is measured using the time-of-flight measurement method. The pump laser is a mode-locked pulse laser with a pulse width of about 15 ps and a repetition rate of 18 MHz. The wavelength is 1552.52 nm, which is over 240 nm away from $\lambda_{ZGVD}$ of SMF-28 (~ 1310 nm), and a variable attenuator adjusts the input pump power. A set of 100-GHz commercial dense wavelength division multiplexing (DWDM) filters with a full-width-at-half-maximum bandwidth of about 0.6 nm remove amplified spontaneous emission noise photons from the pump laser as a band-pass filter (BPF). 200-GHz DWDM filters with a full-width-at-half-maximum bandwidth of about 1.2 nm block the pump light as a notch filter. Since an SMF with various lengths is placed between the BPF and notch filters, the measured photons are only from the short SMF. A 40-channel 100-GHz DWDM module spectrally separates signal and idler photons. The fibers are cooled at liquid nitrogen temperature to reduce the noise photons by spontaneous Raman scattering (SpRS) [8-9]. InGaAs single-photon detectors are used with an efficiency of 5%, a gate width of 3.1 ns, and a dead time of 10 μs. The coincidence window is about 3 ns. To avoid the detector saturation effect due to dead time, we adjust the pump power to maintain single counts of individual detector under 10,000 counts per second. In Fig. 3. (b), the 400-GHz separated channels of the DWDM module from the pump channel are selected for signal and idler photons, respectively, to measure the polarization-entangled states.

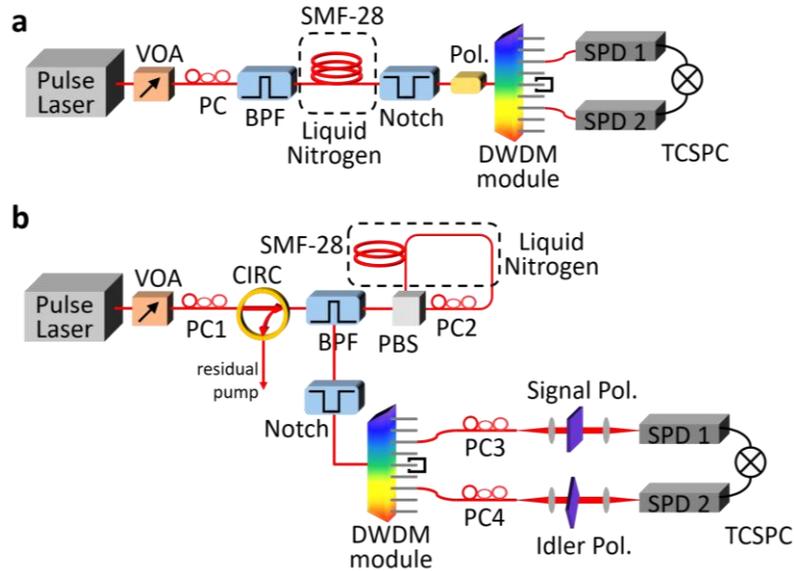

Fig. 3. Experimental set-up (a) for measuring coincidence counts and coincidence-to-accidental ratio and (b) for generating polarization-entangled states. A 15-ps pulse laser is the pump source with a repetition rate of 18 MHz at a center wavelength of 1552.52 nm (ITU grid No. 31). VOA: Variable Optical Attenuator, PC: polarization controller, BPF: band-pass filters for blocking noise photons, Notch: notch filters for pump blocking, Pol.: in-line polarizer, DWDM: 40ch 100-GHz dense wavelength division multiplexer, SPD: InGaAs single photon detector, TCSPC: Time-Correlated Single Photon Counting module, PBS: in-line polarization beam splitter, Signal (Idler) Pol.: free-space polarizer with a rotational stage for signal (idler) photons.

## 4. Results

### 4.1 Coincidence histogram

Since generated photons have a strong temporal correlation, coincidence counts from true photon pairs should be larger than accidental coincident counts. Fig. 4. shows the measured coincidence count histograms for fiber lengths of 3.8 m, 11.4 m, 31.5 m, and 308 m by accumulating the histogram for 10 minutes. The signal and idler filters are 400-GHz separated from pump. The peaks at 0 ns represent the coincidence counts from mostly SFWM photons, and the peaks at about 55.56 ns (inset (i)) and 111 ns (inset (ii)) are the accidental counts by noise photons generated by SpRS and uncorrelated photon pairs. As the pump laser is a pulse laser and the detectors operate in the gating mode, the accidentals do not spread out evenly in time. The peak-to-peak time difference is about 55.56 ns corresponding to the pump repetition rate (18 MHz). For the 3.8-m and 11.4-m fiber whose bandwidth is larger than the filter separation, the coincidence counts are much larger than the accidental coincidence counts (Fig. 4. (a) and (b)). After 11.4 m, the coincidence counts decrease with increasing the fiber length (Fig. 4. (c) and (d)), while the accidental count peaks at about 55.56 ns and 111 ns show no significant change. Due to the narrow bandwidth of the long fibers, C-band pair generation had not previously been observed in SMF-28 fiber, and the results in Fig. 4. (a)-(c) indicate the first experimental implementation of C-band photon-pair generation in conventional SMF-28 fiber.

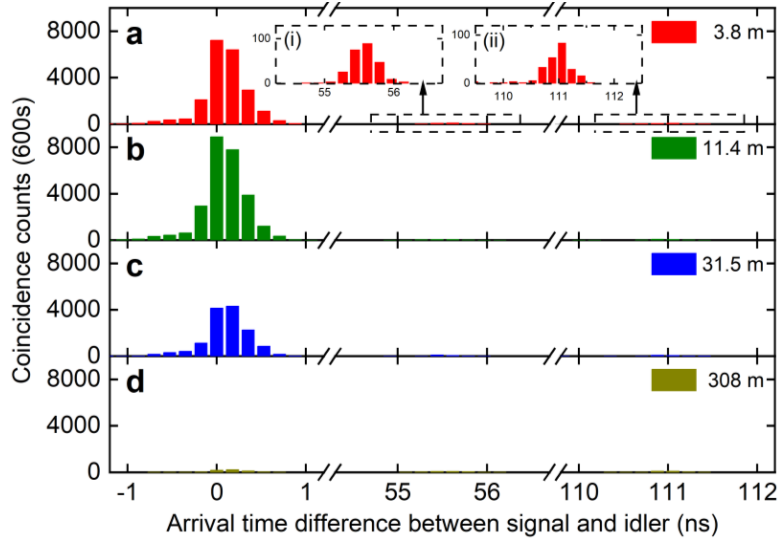

Fig. 4. Measured coincidence histograms for several fiber lengths. The lengths of the fiber under test are (a) 3.8 m, (b) 11.4 m, (c) 31.5 m, and (d) 308 m. Insets (i) and (ii) of (a) show the magnified accidental count peaks, and the time interval of peaks (~55.56 ns) are equivalent to the pump pulse repetition rate (18 MHz). The histogram of accidental counts for fiber lengths of 11.4, 31.5, and 308 m are similar in number and width to those for the 3.8 m fiber and are not shown. The histograms are measured at the consistent single count rate of about 3000 cps, and the accumulation time is 10 minutes. The time resolution of the TCSPC is 176 ps. The signal and idler filters are 400-GHz separated from pump.

## 4.2 Photon-pair generation rate

In order to investigate the performance of SMF-28 fiber as a photon-pair source, we measure the true coincidence counts (coincidence ($C_c$) – accidental ($C_a$)). Fig. 5. (a). show the true coincidence counts for various pump peak powers. The frequency difference between the signal (and idler) and the pump photons is 400 GHz, and each data point is accumulated for 600 seconds. The vertical error bars represent the shot noise, assuming a Poisson distribution of photons. The SFWM process produces photon pairs with a quadratic dependence on the pump power. The dashed lines in Fig. 5. (a). have a slope of 2, representing the quadratic model on the pump power and match well with the data for 3.8-m (red), 11.4-m (green), and 31.5-m (blue) fibers. The quadratic model for 308-m (dark yellow dashed line) fiber, however, does barely fit the data due to the low CAR values (~2) as shown in Fig. 5. (c). Therefore, the coincidence counts for the 308-m fiber are almost equally contributed by SFWM photon pairs and noise photons. Fig. 5. (b). show the true coincidence counts for various fiber lengths. As in Eq. (1), the SFWM process produces photon pairs with a quadratic dependence on the pump power and fiber length. The dashed lines in Fig. 5. (b). have a slope of 2, representing the quadratic model on the fiber length and match well with the data up to 11.4 m, indicating that true coincidence counts increase with the fiber length up to 11.4 m but decrease gradually for long fibers. The reason for noise photons to be dominant in long fibers is the SFWM bandwidth becomes narrow as shown in Fig. 2. (a)., and this is why C-band photon pairs have not been observed in SMF-28 fiber. We additionally measure the coincidence-to-accidental ratio (CAR, $C_c/C_a$), comparing it with the behaviors of true coincidences. As shown in Fig. 5. (c)., the measured CAR values reach above 100 for 3.8-m and 11.4-m fibers and close to 100 for the 31.5-m fiber, but the coincidence counts contributed by noise photons in the 308-m fiber are comparable with that by the SFWM photon pairs having a CAR value of ~2. As shown in Fig. 5. (d)., the CAR values with a pump peak power of 3 W are more than 100 for up to 11.4-m fiber but drop near 2 for 308-m fiber due to the decrease of SFWM bandwidth as in Fig. 2. (a). Therefore, we can conclude that a short SMF-28 fiber can be used as a photon-pair light source with a good signal-to-noise ratio.

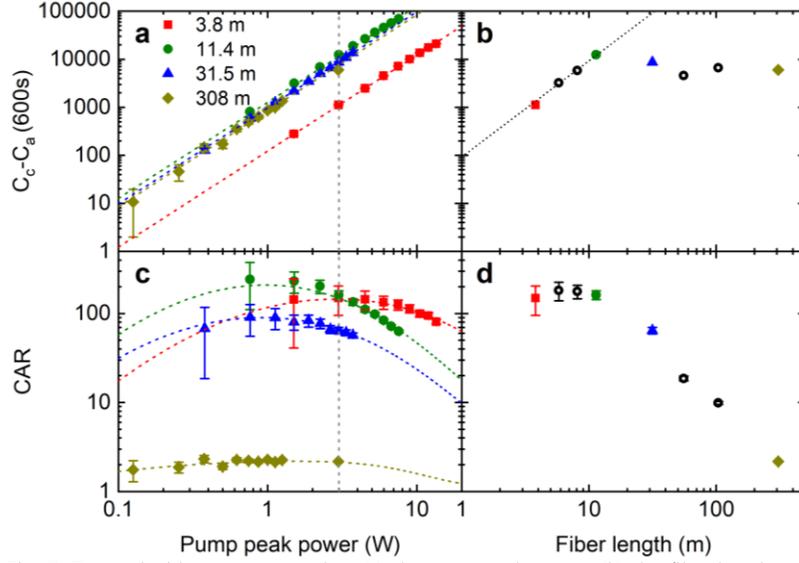

Fig. 5. True coincidence counts against (a) the pump peak power, (b) the fiber length; and coincidence-to-accidental ratio against (c) the pump peak power, (d) the fiber length with 400-GHz filter separation from pump. The pump peak power in (b) and (d) is fixed at about 3 W (grey vertical dashed line in (a) and (c)). The accumulation time of each data point is 600 seconds. Vertical error bars represent the shot noise, assuming a Poisson distribution of photons. The horizontal and vertical axes are plotted on logarithmic scales.

In addition, we investigate the bandwidth of photon-pair generation coefficient ($\mu_p$) depending on fiber length. The PGR bandwidth is examined with various DWDM filter sets in Fig. 3. (a). Using the true coincidence count results in Fig. 5. (a) and (b), $\mu_p$ can be estimated as the total coincidence counts, $C_c$, is given by $C_c = \mu_p \eta_s \eta_i L^2 P_p^2 T + C_a$, where $\eta_s$ ($\eta_i$) is the product of the total transmittance and the detection efficiency of detector for signal (idler) channel and $C_a$ is the accidental coincidence. $T$ is total accumulation time. Therefore, $\mu_p$ can be obtained by using 2nd-order term of quadratic fitting of $C_c - C_a$. The total coincidence detection efficiency $\eta_s \times \eta_i$ is about 0.001 in our system (transmission of each photon: ~ 60%, detection efficiency: ~5%). Fig. 6. shows the theoretically predicted and experimentally measured $\mu_p$ for 3.8-m, 11.4-m, 31.5-m, and 308-m fibers against the frequency detuning ($\Delta \nu$) between the signal (or idler) and pump. The pump peak power is fixed at about 3 W. As shown in Fig. 6, the shorter the fiber length is, the broader the bandwidth is. The theoretically predicted $\mu_p$ in Fig. 2. (a), plotted on a logarithmic scale, are shown with the measured data. The measured results shows a similar tendency with the calculation. In the case of 308-m fiber, $C_c - C_a$ is unable to be fitted for the 800-GHz and 1000-GHz separations due to the low CAR values caused by the narrow bandwidth, as shown in Fig. 5. (c). In the case of 11.4-m fiber, the value of $\mu_p$ at 1000 GHz is larger than that at 800 GHz as the rate at 1000 GHz corresponds to the second peak of $sinc^2$-function. To obtain 3 kHz single count rate, about 13.5-W pump peak power is needed for 3.8-m fiber while 0.75-W is needed for 308-m fiber, $P_p \times L$, however, is about 51 W · m for 3.8-m and 230 W · m for 308-m, including SpRS noise photon. The factor $P_p \times L$ is similar to our previous result ($P_p \times L \sim 65$ W · m for to obtain 3-kHz single count rate) which is used 500-m DSF with the condition of $\lambda_p \sim \lambda_{ZGVD}$ [33]. The measured maximum $\mu_p$ (~ 15 pairs/W$^2$m$^2$s) is also similar with our previous results using a 500-m DSF fiber (~ 20 pairs/W$^2$m$^2$s) [33], as the materials and geometries of SMF and DSF are almost same.

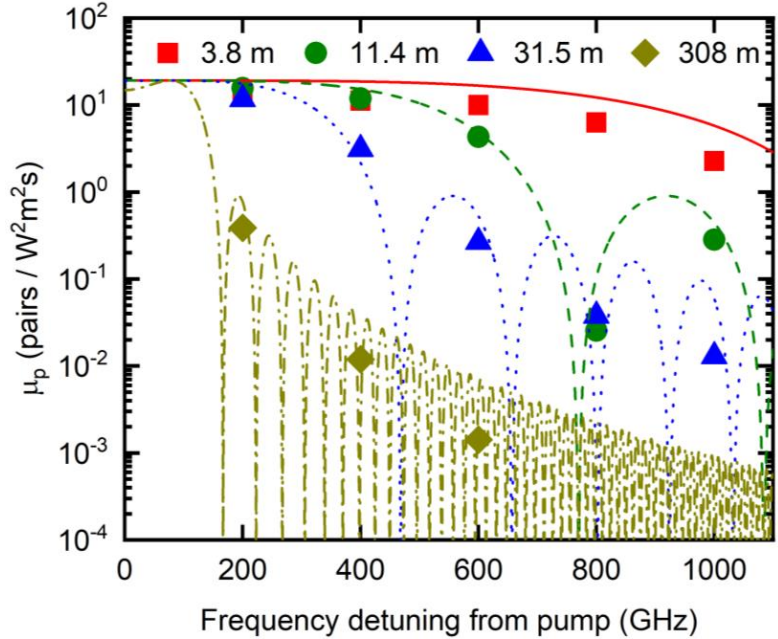

Fig. 6. Calculated (lines) and measured (symbols) pair-generation coefficients $\mu_p$ against frequency separation between signal (or idler) and pump for 3.8 m (red, solid, square), 11.4 m (green, dashed, circle), 31.5 m (blue, dotted, triangle), and 308 m (dark yellow, dash-dotted, diamond) fibers. The vertical axis is plotted on a logarithmic scale, and the horizontal axis is on a linear scale.

### 4.3 Polarization entanglement

Finally, we create polarization-entangled states and measure the entanglement properties. In Fig. 3. (b)., the 400-GHz separated channels of the DWDM module from the pump channel are selected for signal and idler photons, respectively, to measure the polarization-entangled states. An 11.4-m SMF-28 fiber is used. The polarization-entangled states are generated with the well-known Sagnac loop method, as shown in Fig. 3. (b) [34]. Pump light with the diagonal polarization state splits into two orthogonally-polarized pulses ($|H\rangle_p$ and $|V\rangle_p$) with equal power from a fiber-based polarization beam splitter (PBS). $|H\rangle$ and $|V\rangle$ represent the horizontal and vertical polarization states, respectively. The pump pulse with the $|H\rangle(|V\rangle)$ state can generate a photon pair having $|H\rangle_s|H\rangle_i$ ($|V\rangle_s|V\rangle_i$) polarization states through an SMF-28 fiber. The polarization directions rotate 90 degrees by a polarization controller (PC2 in Fig. 3. (b)). When the SFWM photons recombine at the PBS, the SFWM photons are emitted into the PBS port, where the pump beam entered in. Then, the polarization states of the SFWM photons after the PBS become the polarization-entangled states of $|H\rangle_s|H\rangle_i \pm |V\rangle_s|V\rangle_i$. The pump light and polarization-entangled photons are separated by a band-pass filter, which is a commercial DWDM filter with two output ports. The pump light passes through the transmission port, and the SFWM photons reflect into the reflection port. A notch filter having an isolation of over 150 dB suppresses reflected pump light. The SFWM photons split into the two ports of a DWDM module, and then the signal and idler photons pass through the free-space linear polarizer, as seen in Fig. 3. (b). We separately rotate the signal and idler polarizer angles, measuring the coincidence counts between the signal and idler detectors.

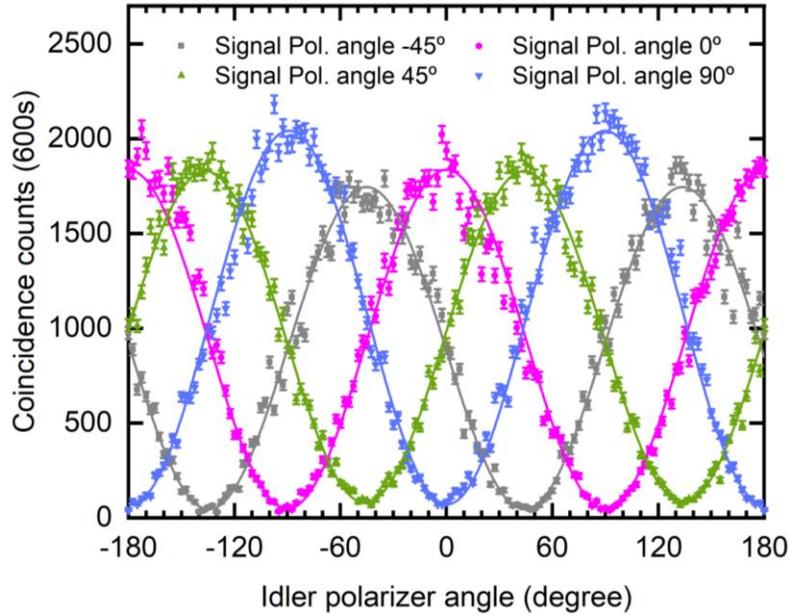

Fig. 7. Measured coincidence counts for several signal polarizer angles against idler polarizer angles. The signal polarizer angles are at -45° (grey), 0° (red), 45° (green), or 135° (blue). All visibilities are over 90% without subtracting accidental coincidence counts. The accumulation time of each point is 600 seconds, and the vertical error bars represent the shot noise assuming a Poisson distribution of photons.

The angle of the signal polarizer ($\theta_1$) is fixed at 0°, and that of the idler polarizer ($\theta_2$) is rotated from -180° to 180° while measuring the coincidence counts by every 2.5°. Then we repeat the measurement for various $\theta_1$ angles of -45°, 45°, and 90°. Fig. 7 shows the measured coincidence count fringes for $\theta_1$ = -45°, 0°, 45°, and 90°, and the accumulation time for each point is 600 seconds. The visibilities of each pattern are 95.4%, 91.3%, 92.5%, and 94.7%, respectively, and the averaged visibility is 94.2% which is equivalent to a Clauser–Horne–Shimony–Holt (CHSH) parameter of $S = 2.664 \pm 0.028$, indicating that the polarization-entangled states using an SMF pair source violate the Bell's inequality. The S value is confirmed by an additional measurement with the 16 combinations of the polarizer settings ($\theta_1$ = -45°, 0°, 45°, 90°; $\theta_2$ = -22.5°, 22.5°, 67.5°, 112.5°) for the CHSH inequality [35, 36]. The accumulation time is 600 seconds for each measurement. We achieve an $S$ value of $S = 2.659 \pm 0.094$ without subtracting the accidental coincident counts, which is very similar to the $S$ value by extracting from the visibility, and the error is estimated from the shot noise assuming a Poisson distribution of photons. The results show that the short fiber with broad bandwidth due to phase mismatching has excellent potentials as a practical and straightforward quantum light source and will contribute significantly to the development of quantum optics experiments.

## 5. Conclusions

In conclusion, we present the method using short SMF-28 fibers as a C-band photon-pair source and show that substantial numbers of photon-pairs can be generated even in the condition of $\lambda_p - \lambda_{ZGVD} > 240$ nm, and the quantum correlation property and entanglement of this photon-pair are sufficiently equivalent to the previously reported results and good for quantum information process. In this study, we employ SMF fiber to verify our method, but any medium with sufficiently large nonlinear Kerr coefficient can be a useful photon-pair source regardless its dispersion properties and pump wavelength. For instance, the geometry of silicon waveguides is usually designed to provide phase-matching, but according to the present study, any short waveguide can create correlated pairs due to small amount of phase mismatch parameter. Further research is on progress.

In addition, we caution that photon pairs from the short sections of single-mode fibers in an experiment setup can be a noise source, as short single-mode fibers are used in almost all fiber-based quantum optical systems as a transmission channel or connection channel between devices in the system. If a strong pump is not filtered out as soon as possible, this pump will make spectrally broad photon pairs in short SMFs used as a transmittance channel. These unexpected photons can degrade the quality of photon states and quantum optical systems.

**Funding.** National Research Foundation of Korea (NRF-2019M3E4A1079780); Korea Institute of Science and Technology's Open Research Program (2E30620-20-052); Institute for Information & communications Technology Promotion(IITP) grant funded by the Korea government(MSIT) (No. 2020-0-00947). RWB acknowledges support through the Natural Sciences and Engineering Research Council of Canada, the Canada Research Chairs program, and the Canada First Research Excellence Fund award on Transformative Quantum Technologies.

**Acknowledgment.** We thank Dr. Yoon-Ho Kim for valuable discussions.